\documentclass[12pt]{article}

\usepackage[all]{xy}
\usepackage{pict2e}
\author{Yu.~M.~Zinoviev
       \thanks{E-mail address: Yurii.Zinoviev@ihep.ru} \\[0.5cm]
        {\it Institute for High Energy Physics} \\
        {\it of National Research Center "Kurchatov Institute"} \\
        {\it Protvino, Moscow Region, 142280, Russia}}

\title{On the Fradkin-Vasiliev formalism in $d=4$}

\date{}

\begin{document}

\maketitle

\begin{abstract}
Here we provide a short review on the so-called Fradkin-Vasiliev
formalism for the construction of higher spin cubic interactions.
Initially it was formulated for the massless fields only, but later on
it was extended to the arbitrary collections of massive and massless
fields.
\end{abstract}

\thispagestyle{empty}
\newpage
\setcounter{page}{1}

\section{Introduction}

Our aim here is to give a short review of the so-called
Fradkin-Vasiliev formalism for the construction of higher spin cubic
interactions. Initially it was formulated for the massless fields
only, but later on its application was extended to arbitrary systems
of massless and massive fields. Such application strongly depends on
the set of fields one tries to investigate and so we divide our review
into four parts.
\begin{enumerate}
\item Three massless fields.
\item One massless field and two massive one. This case in many
respects appears to be special and deserves separate consideration.
In particular, this case contains two physically important tasks:
interaction of massive higher spin fields with gravity and interaction
of massive higher spin supermultiplets with supergravity.
\item Three massive fields.
\item Partially massless fields. These fields are very often
considered as some exotic case, but in many respects they occupy an
intermediate place between massless and massive fields so their
investigation can give  us some useful experience.
\end{enumerate}
We restrict ourselves with $d=4$ not only by technical reasons (such
that the usage of multispinor formalism) but mainly due to the fact
that classification of cubic vertices in $d=4$ drastically differs
from that in $d > 4$.

\section{Massless fields}

Initially, the Fradkin-Vasiliev formalism was proposed for the
construction of the massless higher spin field interactions
\cite{FV87,FV87a} (see also \cite{Vas11,BPS12})). It is based on the
so-called frame-like gauge invariant formalism constructed by Vasiliev
\cite{Vas80,LV88,Vas88} as a generalization to higher spins of the
well known frame-like formalism in gravity.

\subsection{Frame-like formalism for the massless fields}

In this formalism each massless spin-$s$ field is described by the set
of one-forms $\Omega^{\alpha(s-1+m)\dot\alpha(s-1-m)}$, $0 \le |m| \le
s-1$ having some number of dotted and undotted spinor indices 
$\alpha, \dot\alpha = 1,2,$. The case $m=0$ corresponds to the
physical field, $m=1$ --- to the auxiliary fields (similar to the
Lorentz connection in gravity), while all other fields are called
extra ones. Note that the extra fields do not enter the free
Lagrangian but play an important role in the interactions. The reasons
for the appearance of the extra fields are twofold. From one hand,
they allows one to realize a complete set of gauge symmetries. Indeed,
each extra field has its own gauge transformations and its own gauge
invariant two-form (curvature):
\begin{eqnarray*}
\delta \Omega^{\alpha(s-1+m)\dot\alpha(s-1-m)} &=& D 
\eta^{\alpha(s-1+m)\dot\alpha(s-1-m)} + \dots, \\
{\cal R}^{\alpha(s-1+m)\dot\alpha(s-1-m)} &=& D
\Omega^{\alpha(s-1+m)\dot\alpha(s-1-m)} + \dots
\end{eqnarray*}
From the other hand, it is well known that higher spins implies higher
derivatives in the interactions. But contrary to the metric-like
formalism, working with forms one can not have as many explicit
derivatives as one likes and the extra fields serve as an effective
way to describe higher derivatives of the physical field (see below).

In the metric-like formalism one considers statements "on equations of
motion" and "on-shell" as equivalent ones.  But the extra fields do
not enter the free Lagrangian so we have to define  what is 
"on-shell". The answer comes from the so-called First on Shell Theorem
(see \cite{BUV21} and references therein):
\begin{eqnarray}
0 &\approx& D \Omega^{\alpha(s-1+m)\dot\alpha(s-1-m)} + 
e_\beta{}^{\dot\alpha} \Omega^{\alpha(s-1+m)\beta\dot\alpha(s-2-m)} 
+ O(\lambda^2) \nonumber \\
0 &\approx& {\cal R}^{\alpha(2s-2)} - E_{\beta(2)} 
W^{\alpha(2s-2)\beta(2)} \\
0 &\approx& D W^{\alpha(2s+k)\dot\alpha(k)} + e_{\beta\dot\beta}
W^{\alpha(2s+k)\beta\dot\alpha(k)\dot\beta} + \lambda^2
e^{\alpha\dot\alpha} W^{\alpha(2s+k-1)\dot\alpha(k-1)} \nonumber
\end{eqnarray}
Thus all the gauge invariant curvatures except highest ones (which
have only dotted or only undotted indices) vanish on-shell. 
In-particular, it means that each extra field is equivalent to a
derivative of the previous one and as a result to some higher
derivative of the physical field. In turn, the highest curvatures can
be expressed in terms of gauge invariant zero-form (generalization of
the Weyl tensor in gravity). This zero-form is just the first
representative of the infinite set of gauge invariant zero-forms
satisfying so-called unfolded equations and describing all gauge
invariant combinations of the higher derivatives of physical field
which do not vanish on-shell. 

One more important property of the frame-like formalism is that using
these gauge invariant curvatures one can rewrite the free Lagrangian
in the explicitly gauge invariant form:
\begin{equation}
{\cal L}_0 = \sum_{m=1}^{s-1} c_m 
{\cal R}_{\alpha(s-1+m)\dot\alpha(s-1-m)}
{\cal R}^{\alpha(s-1-m)\dot\alpha(s-1-m)} + h.c.
\end{equation}
where coefficients $c_n$ are determined by the so-called extra field
decoupling condition.

\subsection{Cubic vertices}

A classification of the cubic vertices in $d=4$ has been constructed
using a light-cone formalism in \cite{Met18a,Met22} (see
\cite{Met05,Met07b,Met12} for arbitrary dimensions). For the massless
fields it contains two types of vertices\footnote{Relay there exist a
third type but such vertices can not be reproduced in the Lorentz
covariant formalism.} which differ by the number of derivatives (here
and in what follows we assume that $s_1 \ge s_2 \ge s_3$):
\begin{itemize}
\item Type I: $n = s_1 + s_2 + s_3$
\item Type II: $n = s_1 + s_2 - s_3$
\end{itemize}
Let us consider type I vertices. Naively, we have enough derivatives
to construct trivially gauge invariant expressions:
\begin{equation}
{\cal L}_1 \sim W^{\alpha(\hat{s}_2)\beta(\hat{s}_3)} 
W^{\gamma(\hat{s}_1)}{}_{\beta(\hat{s}_3)}
W_{\alpha(\hat{s}_2)\gamma(\hat{s}_1)}
\end{equation}
The only restriction comes from the requirement that all spinor
indices must be contracted for the vertex to be Lorentz invariant.
Taking into account that the total number of indices are determined by
spin of the field:
$$
\hat{s}_2 + \hat{s}_3 = 2s_1, \qquad
\hat{s}_1 + \hat{s}_3 = 2s_2, \qquad
\hat{s}_1 + \hat{s}_2 = 2s_3
$$
we obtain
\begin{equation}
\hat{s}_1 = s_2 + s_3 - s_1, \qquad
\hat{s}_2 = s_1 + s_3 - s_2, \qquad
\hat{s}_3 = s_1 + s_2 - s_3
\end{equation}
But the number of indices can not be negative and this means that for
such vertices to exist the spins $s_{1,2,3}$ must satisfy a so-called
triangular inequality. 

Now let us turn to the type II vertices which require non-trivial
corrections to the gauge transformations. In the constructive approach
to the cubic vertices one has to make two steps. First of all one has
to find cubic terms in the Lagrangian such that their variations under
the initial gauge transformations vanish on the equations of motion.
It means that all variations can be compensated by the appropriate
corrections:
$$
\delta_0 {\cal L}_1|_{e.o.m.} = 0 \quad \Rightarrow \quad
\delta_1 \Omega = \dots
$$
In the frame-like formalism as far as the first step is concerned we
just replace "on the equations of motion" by "on-shell" defined above.
But at the second step we again face the problem related with the fact
that extra fields do not enter the free Lagrangian and so do not have
free equations of motion. The solution has been proposed by Fradkin
and Vasiliev \cite{FV87,FV87a} (see also \cite{Vas11,BPS12}). The main
step of their formalism is to consider the most general consistent 
quadratic deformations for all gauge invariant curvatures
$\hat{\cal R} = {\cal R} + \Delta {\cal R}$ which simultaneously
determines appropriate corrections to the gauge transformations:
$$
\Delta {\cal R} \sim \Omega \Omega \quad \Leftrightarrow \quad
\delta_1 \Omega \sim \Omega \xi 
$$ 
Here the consistency means that all deformed curvatures must transform
covariantly under all gauge transformations:
$$
\delta \hat{\cal R} \sim {\cal R} \xi
$$
Then the interacting Lagrangian is obtained by the replacement of all
curvatures in the free Lagrangian by the deformed ones:
$$
{\cal L} \sim \sum \hat{\cal R} \hat{\cal R} \quad
( + {\cal R} {\cal R} \Omega) 
$$
Note that in space-time dimensions $d > 4$ the general construction
requires also introduction of the so-called abelian terms which are
absent in $d=4$ for the massless fields (however do exist for the
massive fields, see below).

Let us try to apply this formalism for type II vertices
\cite{Vas11,KhZ20a}. Recall that only highest curvatures for each
spin do not vanish on-shell so that any non-trivial deformations must
start with
$$
\Delta {\cal R}_1^{\alpha(2s_1-2)} \sim 
\Omega_2^{\alpha(\hat{s}_3)\beta(\hat{s}_1)}
\Omega_3{}^{\alpha(\hat{s}_2)}{}_{\beta(\hat{s}_1)} + \dots 
$$
It is easy to find that in this case
$$
\hat{s}_1 = s_2 + s_3 - s_1 - 1, \quad
\hat{s}_2 = s_1 + s_3 - s_2 - 1, \quad
\hat{s}_3 = s_1 + s_2 - s_3 - 1
$$
so that the three spins must satisfy a strong triangular inequality.

It is straightforward to check that such procedure generates a lot of
cubic terms with the number of derivatives\footnote{Note that here and
in what follows by the number of derivatives in the vertex we mean the
number of derivatives in terms of the physical fields.} up to
$s_1 + s_2 + s_3 - 2$ which in general mush higher than 
$s_1 + s_2 - s_3$ for the similar flat vertex:
\begin{eqnarray*}
& s_1 + s_2 + s_3 - 2 & \\
& s_1 + s_2 + s_3 - 3 & \\
 & \cdots & \\
& s_1 + s_2 - s_3 + 1 & \\
& s_1 + s_2 - s_3 & \\
& \cdots & 
\end{eqnarray*}
Moreover, the gauge invariance requires that the terms with the
highest number of derivatives combine into total derivative. In some
sense one can say \cite{Vas11} that non-trivial cubic vertex in
$AdS_4$ appeared as a deformation of the trivial flat one. But in our
work \cite{KhZ20a} we showed that all the terms having the number of
derivatives greater than $s_1+s_2-s_3$ vanish on-shell or combine into
total derivative. So finally we can state that there is a one to one
correspondence between flat and $AdS_4$ cubic vertices in agreement
with the general results in \cite{Met18a}. 
For the case when all three spins are different, the flat vertex has a
very simple form \cite{KhZ20a}:
$$
{\cal L}_1 \sim \Omega_1^{\alpha(\hat{s}_2)\dot\alpha(\hat{s}_3)}
\Omega_2^{\beta(\hat{s}_1)}{}_{\dot\alpha(\hat{s}_3)}
D \Omega_{3,\alpha(\hat{s}_2)\beta(\hat{s}_1)}+ h.c.
$$
Note that the lowest spin enters through its gauge invariant
curvature. This explains in-particular why any attempts in flat case
to construct interactions for massless higher spins with gravity or
massless higher spin supermultiplets with supergravity always lead to
the non-minimal interactions.

For the both types of vertices how to go beyond the triangular
inequality is an open question. Note that such vertices do exist in
the Vasiliev theory, see \cite{Mis17,TV24}.

\subsection{Massless supermultiplets}

Let us consider cubic vertices for massless higher spin $N=1$
supermultiplets. In flat space a classification for such vertices was
given in \cite{Met19a} (see \cite{Met19b} for extended
supersymmetries). Recall that massless $N=1$ supermultiplet contains
one boson and one fermion which differ in spin by 1/2 and one has to
distinguish two types: $(s+1/2,s)$ and $(s,s-1/2)$. Now for the three
supermultiplets $(B_i,F_i)$, $i=1,2,3$ we can construct four
elementary vertices
$$
V_0(B_1,B_2,B_3), \quad V_1(B_1,F_2,F_3), \quad
V_2(F_1,B_2,F_3), \quad V_3(F_1,F_2,B_3).
$$
Note that in terms of physical fields supertransformations look like
(schematically) 
$$
\delta B \sim F \zeta \qquad \delta F \sim d B \zeta
$$
Thus for the variations of different elementary vertices can cancel
each other we must have
\begin{equation} 
N_{BBB} = N_{BFF} + 1
\end{equation}
Using this fact it is straightforward to show that in any combination
of three supermultiplets all four elementary vertices can not have a
correct number of derivatives simultaneously. Indeed, in the
classification of \cite{Met19a} all solutions contain only three of
them.

Now let us turn to the application of the Fradkin-Vasiliev formalism
to such vertices \cite{KhZ20b}. The main idea is very simple. Let us
consider for example curvature deformations for the first
supermultiplet corresponding to all four elementary vertices with
initially four arbitrary coupling constants:
\begin{eqnarray*}
\Delta {\cal R}_1 &=& a_0\Delta {\cal R}_1(\Omega_2,\Omega_3) 
+ a_1 \Delta {\cal R}_1(\Phi_2,\Phi_3), \\ 
\Delta {\cal F}_1 &=& a_2 \Delta {\cal F}_1(\Omega_2,\Phi_3) +
a_3 \Delta {\cal F}_1(\Phi_2,\Omega_3)
\end{eqnarray*}
and require that deformed curvatures transform under the
supertransformations as the undeformed ones. It is clear that the
interacting Lagrangian (the sum of the free Lagrangians where all
curvatures are replaced by the deformed ones) will be invariant under
the supertransformations simply because the initial one was invariant.
As a result, we again obtain a one to one correspondence between flat
and $AdS$ vertices. The main difference is that in $AdS_4$ all four
elementary vertices present but in the flat limit one of the coupling
constants goes to zero in agreement with Metsaev's classification.

\section{One massless and two massive fields}

This case appears to be special and deserves separate consideration.
One of the reasons is that it contains two important tasks:
interaction of massive higher spin fields with gravity and interaction
of massive higher spin supermultiplets with supergravity. The main
base of the Fradkin-Vasiliev formalism is the frame-like gauge
invariant description of the massless higher spin fields. Thus to
extend the application of this formalism to the arbitrary collections
of massless and massive fields we need similar description for the
massive ones.

\subsection{Gauge invariance for massive fields}

As it is well known in $d=4$ massive spin $s$ field in the massless
limit decomposes into the set of massless fields with helicities
$\pm s$, $\pm (s-1)$, $\dots$, $0 (\pm 1/2)$. Thus the frame-like
gauge invariant description for the massive fields
\cite{Zin08b,PV10,KhZ19} (see \cite{Zin01,Met06} for the metric-like
case) is constructed using the appropriate set of massless fields.
The main requirement here is that all gauge symmetries of these
massless fields must be conserved (though modified) to guarantee the
correct number of physical degrees of freedom. A technical but
important property of such construction is that only nearest
helicities are mixed:
$$
\xymatrix{
0 \ar@{<->}[r] & \cdots \ar@{<->}[r] & (k-1) \ar@{<->}[r] & 
k \ar@{<->}[r] & \cdots \ar@{<->}[r] & s }
$$
In-particular, this provides a simple explanation of what are the
so-called partially massless fields and how their description appears
from the general massive case (see below).

Each massless field comes with all their one-forms as well as 
zero-forms. Let us illustrate the result with simple figures.
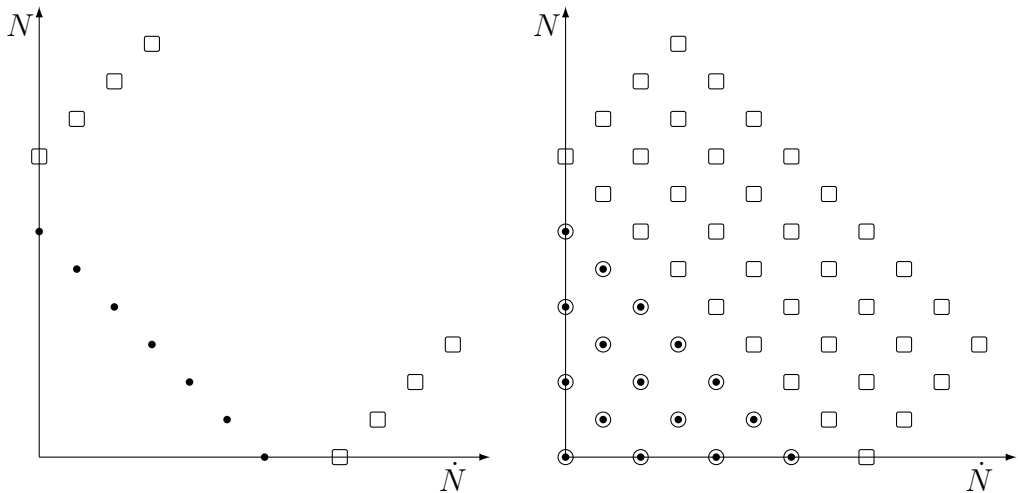
\begin{figure}[htb]
\begin{center}
\begin{picture}(280,140)

\put(10,10){\vector(0,1){120}}
\put(0,120){\makebox(10,10){$N$}}
\put(10,10){\vector(1,0){120}}
\put(115,0){\makebox(10,10){$\dot{N}$}}

\multiput(10,70)(10,-10){7}{\circle*{2}}
\multiput(10,90)(10,10){4}{\oval[0.5](4,4)}
\multiput(90,10)(10,10){4}{\oval[0.5](4,4)}


\put(150,10){\vector(0,1){120}}
\put(140,120){\makebox(10,10){$N$}}
\put(150,10){\vector(1,0){120}}
\put(255,0){\makebox(10,10){$\dot{N}$}}

\put(150,10){\circle*{2}}
\put(150,10){\circle{4}}

\multiput(150,30)(10,-10){3}{\circle*{2}}
\multiput(150,30)(10,-10){3}{\circle{4}}

\multiput(150,50)(10,-10){5}{\circle*{2}}
\multiput(150,50)(10,-10){5}{\circle{4}}

\multiput(150,70)(10,-10){7}{\circle*{2}}
\multiput(150,70)(10,-10){7}{\circle{4}}

\multiput(150,90)(10,10){4}{\oval[0.5](4,4)}
\multiput(160,80)(10,10){4}{\oval[0.5](4,4)}
\multiput(170,70)(10,10){4}{\oval[0.5](4,4)}
\multiput(180,60)(10,10){4}{\oval[0.5](4,4)}
\multiput(190,50)(10,10){4}{\oval[0.5](4,4)}
\multiput(200,40)(10,10){4}{\oval[0.5](4,4)}
\multiput(210,30)(10,10){4}{\oval[0.5](4,4)}
\multiput(220,20)(10,10){4}{\oval[0.5](4,4)}
\multiput(230,10)(10,10){4}{\oval[0.5](4,4)}

\end{picture}
\end{center}
\caption{Massless vs massive case. }
\end{figure}
The left figure is an example of massless field. Here $N$, $\dot{N}$
are the number of dotted and undotted indices, dots stand for the
one-forms, while squares denote gauge invariant zero-forms. The right
figure provides an example of the massive field. Here the circles
denote zero-forms which were gauge invariant in the massless case but
now they are Stueckelberg zero-forms with non-homogeneous
transformation laws typical for the theories with spontaneous symmetry
breaking. It is easy to see that we have one to one correspondence
between gauge and Stueckelberg fields and it is quite natural because
in the massive case all gauge symmetries must be spontaneously broken.

Moreover, besides the gauge invariant two forms ${\cal R}$
corresponding to each gauge field, now we have gauge invariant 
one-form ${\cal B}$ (which we also call curvatures) corresponding to
each Stueckelberg field. Moreover, the free Lagrangian for the massive
field can be rewritten in the explicitly gauge invariant form
$$
{\cal L}_0 \sim \sum [ {\cal R} {\cal R} + {\cal R} {\cal B} + 
{\cal B} {\cal B}]
$$
which is essential property for application of the Fradkin-Vasiliev
formalism.

An important question: what is "on-shell" now. Our analysis shows:
$$
{\cal R}^{\alpha(2s-2)} + h.c., \qquad 
{\cal B}^{\alpha(2s-2-k),\dot\alpha(k)}, \qquad 0 \le k \le 2s-2
$$
so that we have $2s+1$ non-zero curvatures as it should be. Note also
that due to the presence of gauge invariant one-forms abelian vertices
do exist even in $d=4$.

\subsection{Cubic vertices}

In the classification of the cubic vertices with one massless and two
massive fields \cite{Met22} there are two types: with $M_1 \ne M_2$
and $M_1 = M_2$. In general there are more vertices of the second type
and they have less number of derivatives. Trying to investigate such
vertices using gauge invariant formulation of the massive fields one
face a problem related with field redefinitions due to the presence of
Stueckelberg zero-forms. They have non-trivial gauge transformations
and such redefinitions can drastically change the properties of the
vertex (though formally vertices connected by field redefinitions
usually considered to be equivalent, when the redefinitions contain
higher derivatives are such vertices physically equivalent is an open
question). In all cases investigated using the Fradkin-Vasiliev
formalism it appeared that all deformations can be transformed into
abelian form
$$
\Delta {\cal R} \sim {\cal B} \Phi \Rightarrow \delta \Phi \sim 
{\cal B} \xi
$$
so that corrections to the gauge transformations are non-trivial but
all their commutators are zero and the theory remains to be abelian
(hence the name).

It is still has to be proven that the statement above holds for all
such vertices. If it happens to be the case, this property can be used
for the classification of vertices because working with abelian
vertices is much simpler than with non-abelian ones. Note also, that
the results obtained in the so-called unitary gauge (when all
Stueckelberg fields are set to zero) do not depend on field
redefinitions thus providing us with the physically important
information.

\subsection{Examples with massless spin 3/2}

Massive $N=1$ supermultiplet contains four fields: 
$(s+1,s+1/2,s+1/2,s)$ or $(s+1/2,s,s,s-1/2)$, but for the cubic
vertices it is enough to consider interaction of massless spin 3/2
gravitino with the pair of boson and fermion which differ in spin by
1/2 (which we call superblock in \cite{BKhSZ19}.

Let us consider a massive superblock $(2,3/2)$ as a first non-trivial
example \cite{Zin24}. The most general ansatz for abelian vertices
looks like (see Appendix A for the frame-like gauge invariant
description of massive spin 2 and 3/2):
\begin{eqnarray}
{\cal L}_a &=& g_1 {\cal R}^{\alpha\beta} {\cal C}_\alpha \Psi_\beta
+ g_2 {\cal B}^{\alpha\beta} {\cal F}_\alpha \Psi_\beta + g_3
\Pi^{\alpha\dot\alpha} {\cal F}_{\dot\alpha} \Psi_\alpha \nonumber \\
 && + f_1 e_\alpha{}^{\dot\alpha} {\cal B}^{\alpha\beta}
{\cal C}_{\dot\alpha} \Psi_\beta + f_2 e^\alpha{}_{\dot\alpha}
{\cal B}^{\dot\alpha\dot\beta} {\cal C}_{\dot\beta} \Psi_\alpha
+ f_3 e^\alpha{}_{\dot\alpha} \Pi^{\beta\dot\alpha}
{\cal C}_\alpha \Psi_\beta + h.c. 
\end{eqnarray}
Note that we do not assume that two masses are equal. The requirement
that this vertex be invariant under the local supertransformations
gives:
\begin{itemize}
\item two solutions which exist for arbitrary masses $M$, $\tilde{M}$
and are equivalent to trivially gauge invariant ones (where gravitino
enters through its gauge invariant curvature $D \Psi_\alpha$);
\item One additional solution which exists for $M^2 = \tilde{M}^2$
only.
\end{itemize}
Moreover, an analysis of the unitary gauge shows that some particular
combination of these vertices reproduces minimal (with no
more than one derivative) vertex.

Similarly \cite{Zin24}, for massive superblock $(5/2,2)$ there exist
three vertices for arbitrary masses (which are also equivalent to  
trivially gauge invariant ones) and just one additional vertex for
equal masses, while for massive superblock $(3,5/2)$ there exist four
vertices for arbitrary masses (which are also equivalent to  
trivially gauge invariant ones) and just one additional vertex for
equal masses. In both cases some particular combination of these
vertices reproduces minimal vertex.

Note also, that in \cite{Zin24}, using an unfolded formulation for the
massive supermultiplets \cite{KhZ20} we managed to find explicit
expressions of the supertransformations for all gauge and
Stueckelberg fields.

\subsection{Examples with massless spin 2}

In \cite{KhZ21} we investigated cubic interactions of massless spin 2
with massive spin 3/2. First of all we explicitly showed that indeed 
by field redefinitions any such vertex can be reduced to the
abelian form. The most general ansatz for abelian vertex looks like:
\begin{equation}
{\cal L}_a = d_1 {\cal F}_\alpha {\cal C}_\beta \omega^{\alpha\beta} +
d_2 {\cal C}_\alpha {\cal C}_{\dot\alpha} e_\beta{}^{\dot\alpha}
\omega^{\alpha\beta} + d_3 {\cal F}_\alpha {\cal C}_{\dot\alpha}
h^{\alpha\dot\alpha} + d_4 {\cal C}_\alpha {\cal C}_\beta 
e^\alpha{}_{\dot\alpha} h^{\beta\dot\alpha} + h.c,
\end{equation}
Here $\omega^{\alpha(2)} + h.c.$ and $h^{\alpha\dot\alpha}$ describe
massless spin 2. It appeared that there are three linearly independent
abelian vertices and only one is equivalent to the trivially gauge
invariant vertex. Moreover, some combination of these vertices
reproduces minimal gravitational interaction which corresponds to the
spontaneously broken $N=1$ supergravity and in the unitary gauge has
the form:
\begin{equation}
{\cal L}_{min} \sim \Phi_\alpha \Phi_{\dot\alpha} 
e_\beta{}^{\dot\alpha} \omega^{\alpha\beta} - D \Phi_\alpha
\Phi_{\dot\alpha} h^{\alpha\dot\alpha} - 2\tilde{M} \Phi_\alpha
\Phi_\beta e^\alpha{}_{\dot\alpha} h^{\beta\dot\alpha} + h.c. 
\end{equation}

In \cite{KhZ21a} we have considered cubic interactions of massless and
massive spin 2 fields. In this case also we have shown that by field
redefinitions any such vertex can be reduced to the abelian form.
It appeared that there exist three independent trivially gauge
invariant vertices and two abelian vertices which can not be reduced
to the trivially gauge invariant ones. Similarly to the previous case 
some particular combination of these vertices reproduces minimal
(with no more than two derivatives) gravitational interaction which
corresponds to the (linearized) bigravity.

In both cases above we restricted ourselves with one massive spin 3/2
or spin 2 fields. It would be instructive to extend such analysis to
the cases with two massive spin 3/2 or spin 2 fields with different
masses.

\section{Three massive fields}

\subsection{General analysis}

The classification of such vertices in $d=4$ \cite{Met22} contains two
different cases
\begin{itemize}
\item critical: $M_1 = M_2 + M_3$
\item non-critical: $M_1 \ne M_2 + M_3$
\end{itemize}
At the same time in \cite{BDGT18} it was shown that in the gauge
invariant formalism not only one always have enough field
redefinitions to bring any such vertex into abelian form, but allowing
for even higher number of derivatives in the field redefinitions it 
is possible to transform the vertex into trivially gauge invariant
form. But in the frame-like formalism the general ansatz for such
vertices (taking into account the $d=4$) looks like
$$
{\cal L}_1 \sim {\cal R} {\cal B} {\cal B} + {\cal B} {\cal B} 
{\cal B}
$$
and its gauge invariance does not depend on masses. Thus we have one
more open question.

\subsection{Examples}

Our first example is massive spin 2 and two massive spin 3/2 with
three different masses. The most general ansatz for the trivially
gauge invariant vertex looks like:
\begin{eqnarray}
{\cal L}_1 &=& g_1 R^{\alpha\beta} {\cal C}_\alpha 
\tilde{\cal C}_\beta
+ g_2 {\cal B}^{\alpha\beta} {\cal F}_\alpha \tilde{\cal C}_\beta
+ g_3 {\cal B}^{\alpha\beta} {\cal C}_\alpha \tilde{\cal F}_\beta
+ g_4 \Pi^{\alpha\dot\alpha} {\cal F}_\alpha 
\tilde{\cal C}_{\dot\alpha} + g_5 \Pi^{\alpha\dot\alpha}
{\cal C}_\alpha \tilde{\cal F}_{\dot\alpha} \nonumber \\
 && + f_1 e_\alpha{}^{\dot\alpha} {\cal B}^{\alpha\beta}
{\cal C}_\beta \tilde{\cal C}_{\dot\alpha} + f_2 
e_\alpha{}^{\dot\alpha} {\cal B}^{\alpha\beta} {\cal C}_{\dot\alpha}
\tilde{\cal C}_\beta + f_3 e^\beta{}_{\dot\alpha}
\Pi^{\alpha\dot\alpha} {\cal C}_\alpha \tilde{\cal C}_\beta + h.c.
\end{eqnarray}
but taking into account a couple of identities
$$
0 = D [ {\cal B}^{\alpha\beta} {\cal C}_\alpha \tilde{\cal C}_\beta ],
\qquad 0 = D [ \Pi^{\alpha\dot\alpha} {\cal C}_\alpha 
\tilde{\cal C}_{\dot\alpha}]
$$
we have six independent terms. In general this Lagrangian contains
terms with up to three derivatives, but in \cite{Zin18}  it was shown
that for arbitrary values of masses there exist minimal vertex with no
more then one derivative. At the same time the massless limit for spin
2 appeared to be singular unless we set masses of the two gravitini
equal. Similarly, the massless limit for one of the gravitini (so that
previously broken supersymmetry is restored) happens to be
non-singular
for equal masses for spin 2 and second gravitino only.

In \cite{KhZ21a} we have investigated a massive spin 2 
self-interaction. We have explicitly checked that by field
redefinitions any such vertex can be reduced to the abelian form. We
have found that there exist for independent abelian vertices and all
of them are equivalent to the trivially gauge invariant ones. Here
also some particular combination of these vertices reproduces minimal 
(having no more that two derivatives) one. An analysis of the unitary
gauge showed that our results completely consistent both with
\cite{BDGT18} as well as with \cite{LMMS21}.

\section{Partially massless fields}

\subsection{Kinematics}

As we have already mentioned, one of the specific properties of the
gauge invariant description for the massive fields is the fact that
only nearest helicities are mixed. In de Sitter space for some special
values of $M^2 \sim \Lambda$ one of the coefficients becomes zero and
the whole system decomposes into two parts (see e.g.
\cite{Zin01,Met06}):
$$
\xymatrix{
0 \ar@{<->}[r] & \cdots \ar@{<->}[r] & (k-1) & 
k \ar@{<->}[r] & \cdots \ar@{<->}[r] & s }
$$
In this, the right part appears to be unitary and corresponds to some
irreducible representation of de Sitter group. Note that in the flat
limit such representation becomes reducible and decomposes into set of
massless fields with helicities $\pm k$, $\pm (k+1)$, $\dots$, 
$\pm s$. In the frame-like formalism such system is described by the
components shown in the left side of Figure 2.
\begin{figure}[htb]
\begin{center}
\begin{picture}(280,140)

\put(10,10){\vector(0,1){120}}
\put(0,120){\makebox(10,10){$N$}}
\put(10,10){\vector(1,0){120}}
\put(115,0){\makebox(10,10){$\dot{N}$}}

\multiput(10,50)(10,-10){5}{\circle*{2}}
\put(10,50){\circle{4}}
\put(50,10){\circle{4}}

\multiput(10,70)(10,-10){7}{\circle*{2}}
\multiput(10,70)(10,-10){2}{\circle{4}}
\multiput(60,20)(10,-10){2}{\circle{4}}

\multiput(10,90)(10,10){4}{\oval[0.5](4,4)}
\multiput(20,80)(10,10){4}{\oval[0.5](4,4)}
\multiput(30,70)(10,10){4}{\oval[0.5](4,4)}

\multiput(70,30)(10,10){4}{\oval[0.5](4,4)}
\multiput(80,20)(10,10){4}{\oval[0.5](4,4)}
\multiput(90,10)(10,10){4}{\oval[0.5](4,4)}


\put(150,10){\vector(0,1){120}}
\put(140,120){\makebox(10,10){$N$}}
\put(150,10){\vector(1,0){120}}
\put(255,0){\makebox(10,10){$\dot{N}$}}

\multiput(160,40)(10,-10){3}{\circle*{2}}

\multiput(170,50)(10,-10){3}{\circle*{2}}

\multiput(150,90)(10,10){4}{\oval[0.5](4,4)}
\multiput(160,80)(10,10){4}{\oval[0.5](4,4)}
\multiput(170,70)(10,10){4}{\oval[0.5](4,4)}

\multiput(210,30)(10,10){4}{\oval[0.5](4,4)}
\multiput(220,20)(10,10){4}{\oval[0.5](4,4)}
\multiput(230,10)(10,10){4}{\oval[0.5](4,4)}

\end{picture}
\end{center}
\caption{Partially massless limit before vs after gauge fixing}
\end{figure}
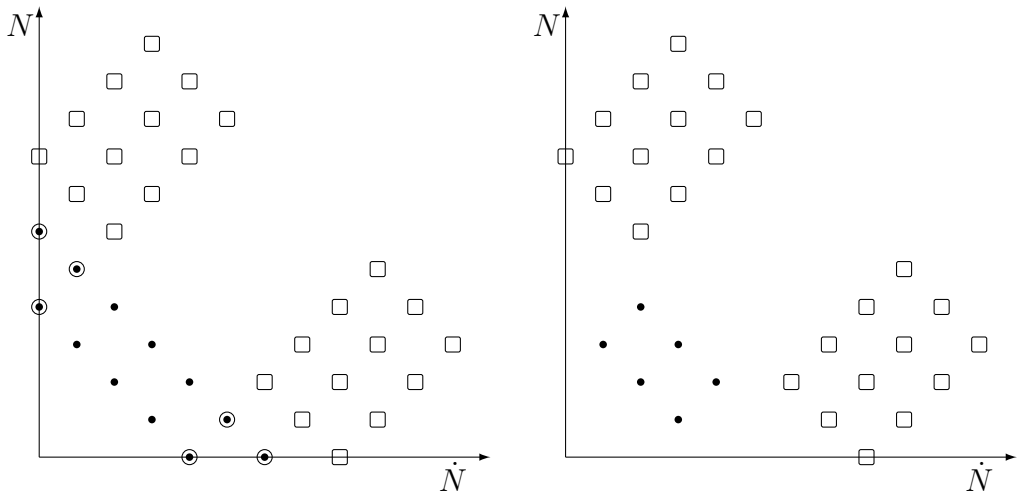
The most important property (besides the fact we have less physical
degrees of freedom) is that now we have some gauge fields without
corresponding Stueckelberg ones. It is also quite natural because in
the partially massless case not all the gauge symmetries are
spontaneously broken (hence the name). 

The right part of Figure 2 illustrates the formalism for the
description of partially massless fields proposed for the bosonic case
by Skvortsov and Vasiliev \cite{SV06} and later on extended on the
fermionic case in \cite{KhZ19}. Such formalism can be considered as
the result of partial gauge fixing when we set all remaining
Stueckelberg fields to zero and solve their equations. We provide an
illustration of such procedure using partially massless spin 5/2 as an
example in Appendix B.

\subsection{Cubic vertices}

Till now there exist just a few explicit examples on the
application of the Fradkin-Vasiliev formalism to partial;y massless
fields (see e.g. \cite{Zin14}), more will appear soon. The situation
with the field redefinitions strongly depends on which formalism one
uses:
\begin{itemize}
\item before gauge fixing: in general we do not have enough to bring
the vertex into abelian form so one has to consider the most general
combination of non-abelian and abelian vertices;
\item after gauge fixing: there are no any ambiguities, formalism
works
exactly as in the massless case.
\end{itemize}
Note however that the two procedures (gauge fixing and switching on
interaction) in general do not commute so the equivalence of the
results is an open question. For the Skvortsov-Vasiliev case it is
easy to find a generalization of the triangular inequality. Let us
denote $l = s-k$ where $k$ is the lowest helicity in the spectrum so
that $l=0$ corresponds to the massless case. Then for the vertex to
exist we must have: 
\begin{eqnarray*}
s_1 - l_1 &<& s_2 - l_2 + s_3 - l_3 \\
l_1 &\le& l_2 + l_3
\end{eqnarray*}
It is easy to see that for the free massless fields $l_{1,2,3}=0$ the
second equation is valid while the first one becomes usual triangular
inequality. Less trivial observation is that the second equation
forbids vertices with two massless and one partially massless fields.
At the algebraic level it means that in any closed system of massless
and partially massless fields, massless ones must correspond to some
closed subalgebra. Note that there exist some candidates for the
infinite dimensional algebra corresponding to such system
\cite{BG13,JM15,BD24}. 

The partially massless fields are unitary in de Sitter space and
non-unitary in anti de Sitter one. Nonetheless, it is interesting
that in $AdS_4$ there exist supermultiplets containing such fields 
\cite{GHR18,BKhSZ19a,BGHR21}.

\appendix

\section{Kinematics}

\subsection{Spin 3/2}

In this case we need only physical fields: one-forms $\Phi^\alpha$, 
$\Phi^{\dot\alpha}$ and Stueckelberg zero-forms $\phi^\alpha$, 
$\phi^{\dot\alpha}$. Their gauge invariant curvatures look like:
\begin{eqnarray}
{\cal F}^\alpha &=& D \Phi^\alpha + \tilde{M} e^\alpha{}_{\dot\alpha}
\Phi^{\dot\alpha} - \frac{a_0}{3} E^\alpha{}_\beta \phi^\beta,
\nonumber \\
{\cal C}^\alpha &=& D \phi^\alpha - a_0 \Phi^\alpha + \tilde{M}
e^\alpha{}_{\dot\alpha} \phi^{\dot\alpha}, 
\end{eqnarray}
where
$$
a_0{}^2 = 6\tilde{M}^2,
$$
These curvatures satisfy the following differential identities, which
play an important role:
\begin{eqnarray}
D {\cal F}^\alpha &=& - \tilde{M} e^\alpha{}_{\dot\alpha}
{\cal F}^{\dot\alpha} - \frac{a_0}{3} E^\alpha{}_\beta
{\cal C}^\beta, \nonumber \\
D {\cal C}^\alpha &=& - a_0 {\cal F}^{\alpha} - \tilde{M}
e^\alpha{}_{\dot\alpha} {\cal C}^{\dot\alpha}. \label{dif_1}
\end{eqnarray}
On-shell we have
\begin{eqnarray}
0 &\approx& D \Phi^\alpha + \tilde{M} e^\alpha{}_{\dot\alpha}
\Phi^{\dot\alpha} - \frac{a_0}{3} E^\alpha{}_\beta \phi^\beta +
E_{\beta(2)} Y^{\alpha\beta(2)}, \nonumber \\
0 &\approx& D \phi^\alpha - a_0 \Phi^\alpha + \tilde{M} 
e^\alpha{}_{\dot\alpha} \phi^{\dot\alpha} + e_{\beta\dot\alpha}
Y^{\alpha\beta\dot\alpha}. 
\end{eqnarray}
Here zero-forms $Y^{\alpha(3)}$ and $Y^{\alpha(2)\dot\alpha}$ are the
first representatives of two infinite sets of gauge invariant 
zero-forms $Y^{\alpha(3+k),\dot\alpha(k)}$ and
$Y^{\alpha(2+k)\dot\alpha(1+k)}$, $k \ge 0$.

\subsection{Massive spin 2}

In this case we need both physical fields: one-forms
$H^{\alpha\dot\alpha}$, $A$ and zero-form
$\varphi$ as well as auxiliary fields: one-forms 
$\Omega^{\alpha(2)}$, $\Omega^{\dot\alpha(2)}$ and zero-forms
$B^{\alpha(2)}$, $B^{\dot\alpha(2)}$ and $\pi^{\alpha\dot\alpha}$.
Gauge invariant curvatures for the physical fields look like:
\begin{eqnarray}
{\cal T}^{\alpha\dot\alpha} &=& D H^{\alpha\dot\alpha} + 
e_\beta{}^{\dot\alpha} \Omega^{\alpha\beta} + e^\alpha{}_{\dot\beta}
\Omega^{\dot\alpha\dot\beta} + M e^{\alpha\dot\alpha} A, \nonumber \\
{\cal A} &=& D A + 2(E_{\alpha(2)} B^{\alpha(2)} + E_{\dot\alpha(2)}
B^{\dot\alpha(2)}) + M e_{\alpha\dot\alpha} H^{\alpha\dot\alpha}, \\
\Phi &=& D \varphi + e_{\alpha\dot\alpha} \pi^{\alpha\dot\alpha}
+ M A, \nonumber
\end{eqnarray}
while for the auxiliary fields
\begin{eqnarray}
{\cal R}^{\alpha(2)} &=& D \Omega^{\alpha(2)} + M E^\alpha{}_\beta
B^{\alpha\beta} + \frac{M^2}{2} e^\alpha{}_{\dot\alpha}
H^{\alpha\dot\alpha} + 2M^2 E^{\alpha(2)} \varphi, \nonumber \\
{\cal B}^{\alpha(2)} &=& D B^{\alpha(2)} + M \Omega^{\alpha(2)} +
\frac{M}{2} e^\alpha{}_{\dot\alpha} \pi^{\alpha\dot\alpha}, \\
\Pi^{\alpha\dot\alpha}  &=& D \pi^{\alpha\dot\alpha} + M 
(e_\beta{}^{\dot\alpha} B^{\alpha\beta} + e^\alpha{}_{\dot\beta}
B^{\dot\alpha\dot\beta}) + M^2 H^{\alpha\dot\alpha} + M^2
e^{\alpha\dot\alpha} \varphi. \nonumber
\end{eqnarray}
Similarly to the well known "zero-torsion condition" in gravity, 
on-shell we put
$$
{\cal T}^{\alpha\dot\alpha} \approx 0, \qquad
{\cal A} \approx 0, \qquad \Phi \approx 0.
$$
and this leaves us with
\begin{eqnarray}
0 &\approx& {\cal R}^{\alpha(2)} + E_{\beta(2)} W^{\alpha(2)\beta(2)},
\nonumber \\
0 &\approx& {\cal B}^{\alpha(2)} + e_{\beta\dot\alpha}
B^{\alpha(2)\beta\dot\alpha}, \\
0 &\approx& \Pi^{\alpha\dot\alpha}  
+ e_{\beta\dot\beta} \pi^{\alpha\beta\dot\alpha\dot\beta}. \nonumber
\end{eqnarray}
Here $W^{\alpha(4)}$, $B^{\alpha(3)\dot\alpha}$ and 
$\pi^{\alpha(2)\dot\alpha(2)}$ are the first representative of the
three infinite sets of gauge invariant zero-forms.

\section{Partially massless spin 5/2} 

Partially massless limit for massive spin 5/2 \cite{KhZ19} leaves us
with  three one-forms $\Omega^{\alpha(3)}$, 
$\Phi^{\alpha(2)\dot\alpha}$, $\Phi^\alpha$ and one zero-form
$C^{\alpha(3)}$. Their gauge invariant curvatures look like:
\begin{eqnarray}
{\cal F}^{\alpha(3)} &=& D \Phi^{\alpha(3)} - \frac{4\rho_2}{5}
E^\alpha{}_\beta C^{\alpha(2)\beta} \nonumber \\
{\cal F}^{\alpha(2)\dot\alpha} &=& D \Phi^{\alpha(2)\dot\alpha} +
e_\beta{}^{\dot\alpha} \Phi^{\alpha(2)\beta} + \rho_1
e^\alpha{}_{\dot\beta} \Phi^{\alpha\dot\alpha\dot\beta} + \rho_2
e^{\alpha\dot\alpha} \Phi^\alpha \nonumber \\
{\cal F}^\alpha &=& D \Phi^\alpha + 3\rho_2 e_{\beta\dot\alpha}
\Phi^{\alpha\beta\dot\alpha} + 3\rho_1 e^\alpha{}_{\dot\alpha}
\Phi^{\dot\alpha} - E_{\beta(2)} C^{\alpha\beta(2)} \\
{\cal C} ^{\alpha(3)} &=& D C^{\alpha(3)} - 6\rho_2 \Phi^{\alpha(3)}
\nonumber
\end{eqnarray}
Here
$$
\rho_1{}^2 = \frac{\lambda^2}{4}, \qquad
\rho_2{}^2 = - \frac{5\lambda^2}{12}
$$
Let us go the the unitary gauge $C^{\alpha(3)} = 0$. To see that we
will still have a correct number of physical degrees of freedom
consider the unfolded equations:
\begin{eqnarray}
0 &=& D \Phi^{\alpha(3)} - \frac{4\rho_2}{5} E^\alpha{}_\beta
C^{\alpha(2)\beta} - E_{\beta(2)} Y^{\alpha(3)\beta(2)} \nonumber \\
0 &=& D \Phi^{\alpha(2)\dot\alpha} + e_\beta{}^{\dot\alpha}
\Phi^{\alpha(2)\beta} + \rho_1 e^\alpha{}_{\dot\beta}
\Phi^{\alpha\dot\alpha\dot\beta} + \rho_2 e^{\alpha\dot\alpha}
\phi^\alpha \nonumber \\
0 &=& D \Phi^\alpha + 3\rho_2 e_{\beta\dot\alpha}
\Phi^{\alpha\beta\dot\alpha} + 3\rho_1 e^\alpha{}_{\dot\alpha}
\Phi^{\dot\alpha} - E_{\beta(2)} C^{\alpha\beta(2)} \\
0 &=& D C^{\alpha(3)} - 6\rho_2 \Phi^{\alpha(3)} 
- e_{\beta\dot\alpha} Y^{\alpha(3)\beta\dot\alpha} \nonumber
\end{eqnarray}
where $Y^{\alpha(5)}$ and $Y^{\alpha(4)\dot\alpha}$ are gauge
invariant zero-forms. At $C^{\alpha(3)} = 0$ the last equation gives
$$
\Phi^{\alpha(3)} = - \frac{1}{6\rho_2} e_{\beta\dot\alpha}
Y^{\alpha(3)\beta\dot\alpha} 
$$
Then the first equation takes the form
$$
0 = e_{\beta\dot\alpha} D Y^{\alpha(3)\beta\dot\alpha}
- 6\alpha_2 E_{\beta(2)} Y^{\alpha(3)\beta(2)}
$$
which is consistent with the unfolded equation for 
$Y^{\alpha(4)\dot\alpha}$. Remaining equations
\begin{eqnarray}
0 &=& D \Phi^{\alpha(2)\dot\alpha} + \rho_1 e^\alpha{}_{\dot\beta}
\Phi^{\alpha\dot\alpha\dot\beta} + \rho_2 e^{\alpha\dot\alpha}
\Phi^\alpha + \frac{2}{3\rho_2} E_{\beta(2)} 
 Y^{\alpha(2)\beta(2)\dot\alpha} \nonumber \\
0 &=& D \Phi^\alpha + 3\rho_2 e_{\beta\dot\alpha}
\Phi^{\alpha\beta\dot\alpha} + 3\rho_1 e^\alpha{}_{\dot\alpha}
\Phi^{\dot\alpha}
\end{eqnarray}
tell us that in the Lagrangian formulation we have just two curvatures
\begin{eqnarray}
{\cal F}^{\alpha(2)\dot\alpha} &=& D \Phi^{\alpha(2)\dot\alpha} 
+ \rho_1 e^\alpha{}_{\dot\beta} \Phi^{\alpha\dot\alpha\dot\beta} +
\rho_2 e^{\alpha\dot\alpha} \Phi^\alpha \nonumber \\
{\cal F}^\alpha &=& D \Phi^\alpha + 3\rho_2 e_{\beta\dot\alpha}
\Phi^{\alpha\beta\dot\alpha} + 3\rho_1 e^\alpha{}_{\dot\alpha}
\Phi^{\dot\alpha}
\end{eqnarray}
Let us stress that now ${\cal F}^{\alpha(2)\dot\alpha} \ne 0$
on-shell. Note also that the free Lagrangian can be written as
\begin{equation}
{\cal L}_0 = a_1 {\cal F}_{\alpha(2)\dot\alpha} 
{\cal F}^{\alpha(2)\dot\alpha} + a_2 {\cal F}_\alpha {\cal F}^\alpha +
h.c.
\end{equation}

\end{document}